\documentclass[fleqn,12pt,twoside]{article}
\usepackage[headings]{espcrc1}

\readRCS
$Id: espcrc1.tex,v 1.2 2004/02/24 11:22:11 spepping Exp $
\usepackage{graphicx}
\usepackage[figuresright]{rotating}

\newcommand{\AmS}{{\protect\the\textfont2
  A\kern-.1667em\lower.5ex\hbox{M}\kern-.125emS}}

\title{Nuclear modification factor for identified hadrons at forward rapidity in Au+Au reactions at 200 GeV}

\author{Radoslaw Karabowicz\address[ZFGM]{M. Smoluchowski Institute of Physics, \\ 
	Jagiellonian University, \\
        Reymonta 4, Krak\'ow 30-059, Poland} %
	 for the BRAHMS\thanks{For the full BRAHMS Collaboration author list and acknowledgment, see appendix 'Collaborations' of this volume.} 
	Collaboration.}
       
\runtitle{Nuclear modification factor in Au+Au reactions at 200 GeV from BRAHMS}
\runauthor{R. Karabowicz}

\begin{document}

\maketitle

\begin{abstract}
Herewith we present the production of identified hadrons in Au~+~Au and p~+~p collisions
at $\sqrt{s_{NN}} = 200 \,{\rm GeV}$ at forward rapidity, $y \approx 3.2$. Suppression of pions
and kaons and enhancement for protons in central Au~+~Au collisions is observed. These results
are found to be very similar in strength to that observed at mid-rapidity.
Furthermore, we see a gradual decrease of the observed suppression towards more peripheral
collisions.
\end{abstract}
\newline

First collisions of gold ions at nucleon-nucleon center of mass energies $130 \,{\rm GeV}$ at RHIC revealed a dramatic 
decrease of pion production at high transverse momentum ($p_T$), as compared to an incoherent sum of pions 
produced in the p~+~p collisions at the same energy~\cite{AdcoAA}. High $p_T$ hadrons are primarily produced from the 
fragmentation of the hard-scattered partons and observed suppression could be either due to initial state 
parton saturation inside the nuclei~\cite{Khar03} or due to final state jet energy degradation~\cite{Bjor86}. 
The crucial test of these different mechanisms has been performed during the third RHIC run when collisions 
between deuterium and gold ions at $\sqrt{s_{NN}} = 200 \,{\rm GeV}$ were investigated. The measurements showed that the high $p_T$ particle 
production from d~+~Au collisions around mid-rapidity is not suppressed~\cite{AdleAA,ArseAA,AdamAA,BackAA}. The absence of this phenomena 
supports the interpretation that the observed suppression in the Au~+~Au collisions is due to the final
interactions.
However it was also observed in the BRAHMS experiment that at forward pseudo-rapidity $\eta = 2.2$ inclusive 
negatively charged hadrons are suppressed in both central Au~+~Au and minimum-bias d~+~Au collisions~\cite{ArseAA,ArseAB},
which might be attributed to the possible existence of the nuclei in the Color Glass Condensate phase~\cite{Khar03}
prior to the collisions. Studying the magnitude of suppression as a function of rapidity and centrality
is essential to help understanding different mechanism responsible for this phenomenon.

To study the in-medium effects on the spectra it is often useful to plot the nuclear modification
factor, which is the ratio of the yield obtained from nucleus-nucleus collisions scaled with the number 
of binary collisions, to the yield from elementary nucleon-nucleon collisions:

\begin{equation}
R_{AA} = \frac {{\rm d}^2N_{AA}/{\rm d}p_T{\rm d}y}{<{N_{coll}>} \times {\rm d}^2N_{NN}/{\rm d}p_T{\rm d}y}.
\end{equation}

In the absence of any nuclear effects the ratio 
should saturate at unity for high $p_T$, where production is dominated by 
hard scatterings and is proportional to the number of binary collisions, $N_{coll}$. Production in the low $p_T$ region
is dominated by soft processes and scales with the number of participants, $N_{part}$, which is {\it circa}
three times smaller for central Au~+~Au collisions than $N_{coll}$.
\section{Results}

\begin{figure}[htb]\centering
\includegraphics[width=35pc]{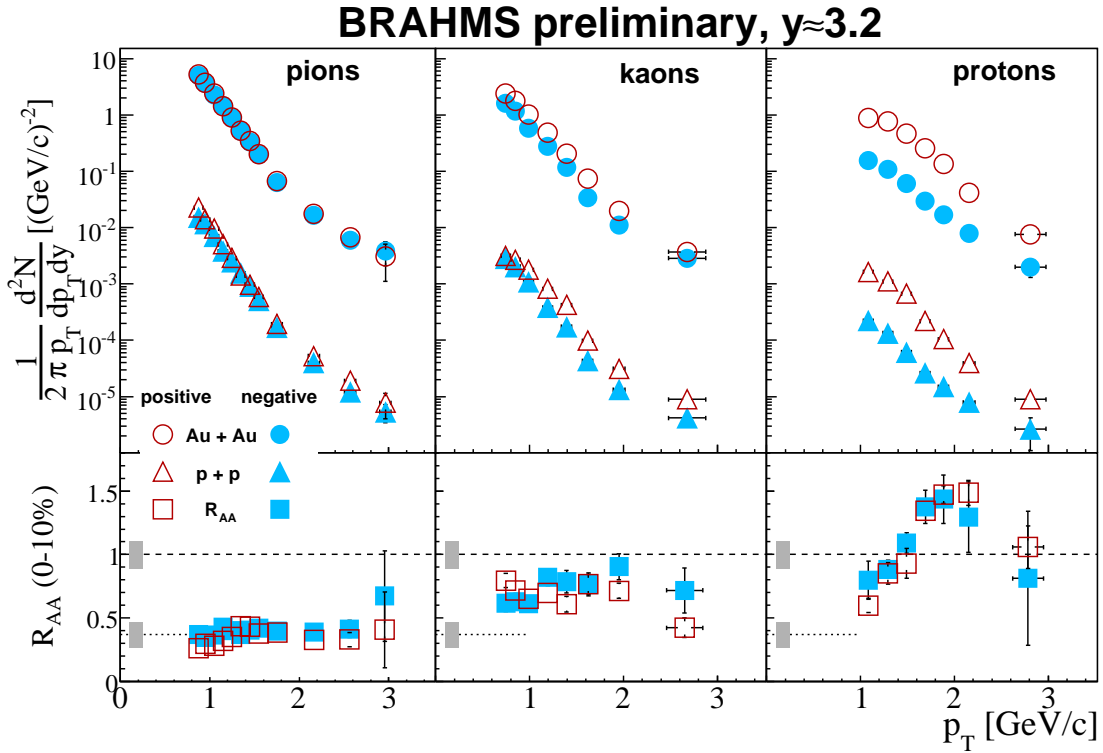}
\vspace{-1cm}
\caption{Top panels show particle transverse momentum spectra for pions (left-hand panel), kaons (middle) 
	 and protons (right-hand) from $0-10\%$ central Au~+~Au (circles) and p~+~p (triangles) 
	 collisions at $200 \,{\rm GeV}$. Bottom panels show nuclear modification factors (squares) for those
	 particles. Solid symbols represent negative particles, while open positive.
	 No correction for decay, absorption or feed-down has been applied to the data. 
	 Only statistical errors are shown.}
\vspace{-0.0cm}
\label{fig:SpectraAndRaa}
\end{figure}

Top panels of Figure~\ref{fig:SpectraAndRaa} shows differential yields per event for identified hadrons 
in Au~+~Au collision at $\sqrt{s_{NN}} = 200 \,{\rm GeV}$ as seen by the BRAHMS experiment 
at rapidity $y \approx 3.2$. Also there we plot the differential
yields for p~+~p collisions. 

In the bottom panels of Figure~\ref{fig:SpectraAndRaa} we plot the $R_{AA}$ for identified particles
at rapidity $y \approx 3.2$ for $0-10\%$ 
central Au~+~Au collisions at $\sqrt{s_{NN}} = 200 \,{\rm GeV}$. Dashed and dotted lines represent the expectation of scaling
with $N_{coll}$ and $N_{part}$, respectively, while the shaded boxes are the 
systematic errors resulting from the uncertainties of $N_{coll}$ and $N_{part}$.
This figure shows suppression of pions (left-hand panel) and 
kaons (middle panel) at values of about 0.4 and 0.7, respectively, similar for both particle signs, and basically independent of $p_T$
in the measured $p_T$ ranges. $R_{AA}$ for protons however, 
plotted in the right-hand panel, exhibit an enhancement peak at $p_T \approx 2 \,{\rm GeV}/c$.
The increase of the calculated nuclear modification factor with increasing mass from pions 
via kaons to protons may be attributed to coalescence or radial flow. 
However explanation of the different shape of $R_{AA}$ for protons requires
invoking particle production mechanisms that depend on the number of quarks, 
such as baryon junction~\cite{Vite02,Topo03} or parton recombination~\cite{Hwa_03}. 

\begin{figure}[htb]\centering
\includegraphics[width=35pc]{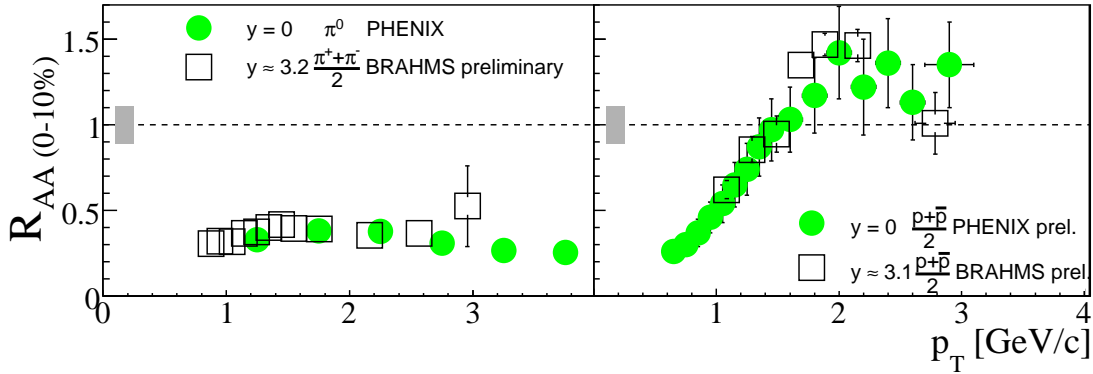}
\vspace{-1cm}
\caption{Comparison of $R_{AA}$ for pions (left-hand panel) and protons (right-hand panel) at mid-rapidity and $y \approx 3.2$.
	 Solid circles are PHENIX data~\cite{AdleAB,Pal_AA} at mid-rapidity, open squares show BRAHMS preliminary data.}
\vspace{-1.0cm}
\label{fig:PhenixRaa}
\end{figure}

In Figure~\ref{fig:PhenixRaa} we compare the $R_{AA}$ calculated for
pions and protons at $y \approx 3.2$ with the ratios obtained by the PHENIX Collaboration at
mid-rapidity. Data show very similar behavior for both rapidities in the $p_T$ range covered
by BRAHMS, which, combined with other BRAHMS
results~\cite{ArseAA}, indicate the flatness of $R_{AA}$ with rapidity for central Au~+~Au collisions.

\begin{figure}[htb]\centering
\includegraphics[width=25pc]{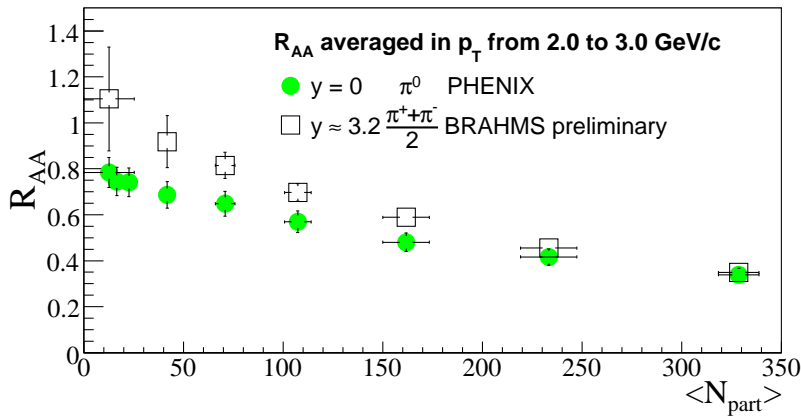}
\vspace{-1cm}
\caption{Change of averaged nuclear modification factor with centrality. 
	 Solid circles are PHENIX data~\cite{AdleAB} at mid-rapidity, open squares show BRAHMS preliminary data.}
\vspace{-0.0cm}
\label{fig:AuAuRaaNpart}
\end{figure}

The magnitude of the suppression can also be studied as a function of the system size.
Averaged $R_{AA}$ for pions in the $p_T$ range from 2.0 to $3.0 \,{\rm GeV}/c$ has been
plotted in Figure~\ref{fig:AuAuRaaNpart} as a function of $<N_{part}>$.
As expected, there is less suppression at peripheral events. 
When compared with the PHENIX data obtained for neutral pions at mid-rapidity
we can see that although for central events $R_{AA}$ seems to be independent of rapidity,
for the peripheral events nuclear modification factor seen at forward rapidity reaches
value close to unity, larger than the PHENIX data. A simple finding
is that similar mechanisms such as parton recombination occur at central collisions at mid- and forward rapidities,
however this similarity breaks down for peripheral events.
\section{Conclusions}

BRAHMS experiment has measured particle distributions for identified hadrons in Au~+~Au and p~+~p collisions
at $\sqrt{s_{NN}} = 200 \,{\rm GeV}$ at forward rapidity of $y \approx 3.2$. Spectra were used to construct the 
nuclear modification factor, which shows suppression for mesons independent of the transverse momenta, 
and enhancement for protons and antiprotons around $p_T \approx 2 \,{\rm GeV}/c$. Systematic difference between mesons and 
baryons, which moreover seems to be independent of rapidity for central events (at least up to rapidity $y \approx 3.2$), 
indicates existence of various mechanisms of particle production. Remarkable independence
of the investigated ratios on rapidity can be attributed to the surface emission picture~\cite{Dain04}, where changes in 
${\rm d}N/{\rm d}\eta$ result in a nearly flat dependence of $R_{AA}$, albeit an increase of $R_{AA}$ is expected
for $y > 3$.

Surprising feature is revealed when comparing nuclear modification factor dependence on centrality
at mid-rapidity and at large rapidity. Observed enhancement of the ratio at $y \approx 3.2$
for the peripheral events may indicate change of the medium properties when going from mid-
to forward rapidity. The conclusion might be drawn that the strongly interacting
matter that quenches particle production at high $p_T$ extends to much higher rapidities in the central events than
in the peripheral events.

\end{document}